\documentclass[preprint,showpacs,preprintnumbers,amsmath,amssymb]{revtex4}

\usepackage{graphicx}
\usepackage{dcolumn}
\usepackage{bm}
\usepackage{graphics}
\usepackage{color}
\usepackage{color}
\usepackage{subfigure}
\usepackage{epsfig}

\begin{document}
\preprint{}

\title{Relativistic models for quasi-elastic neutrino scattering}%

\author{M.C. Mart\'{\i}nez}
\author{P. Lava}
\author{N. Jachowicz}
\author{J. Ryckebusch}
\author{K. Vantournhout }

\affiliation{Department of Subatomic and Radiation Physics, Ghent
  University,
  Proeftuinstraat 86, B-9000 Gent, Belgium}
\author{J.M. Ud\'{\i}as}
\affiliation{Departamento de F\'\i sica At\'omica, Molecular y Nuclear, Facultad de Ciencias F\'\i sicas, 
Universidad Complutense de Madrid, E-28040 Madrid, Spain}
      
\date{\today}

\begin{abstract}
We present quasi-elastic neutrino-nucleus cross sections in the energy range from 150 MeV up to 5 GeV for the target nuclei $^{12}$C and $^{56}$Fe. A relativistic description of the nuclear dynamics and the neutrino-nucleus coupling is adopted. For the treatment of final-state interactions (FSI) we rely on two frameworks succesfully applied to exclusive electron-nucleus scattering: a relativistic optical potential and a relativistic multiple-scattering Glauber approximation. At lower energies, the optical-potential approach is considered to be the optimum choice, whereas at high energies a Glauber approach is more natural. Comparing the results of both calculations, it is found that the Glauber approach yields valid results down to the remarkably small nucleon kinetic energies of 200 MeV. We argue that the nuclear transparencies extracted from $A(e,\, e'p)$ measurements can be used to obtain realistic estimates of the effect of FSI mechanisms on quasi-elastic neutrino-nucleus cross sections. We present two independent relativistic plane-wave impulse approximation (RPWIA) calculations of quasi-elastic neutrino-nucleus cross sections. They agree at the percent level, showing the reliability of the numerical techniques adopted and providing benchmark RPWIA results.
\end{abstract}

\maketitle

\section{Introduction}
Neutrino interactions offer unique opportunities for exploring fundamental questions in different domains of physics. The mass of the neutrino remains one of the greatest puzzles in elementary particle physics. In recent years, a number of positive neutrino oscillation signals  made the claims of non-zero neutrino masses irrefutable \cite{superkam98} and boosted the interest in this issue. Several experiments are running or proposed in order to address intriguing questions in current neutrino physics \cite{freedman}:   What does the neutrino mass hierarchy look like, and what are the values of the oscillation parameters \cite{superkam98}? What is the role of  vacuum and matter-enhanced oscillations ? Are neutrinos representatives of CP-violation in the leptonic sector ? Is the neutrino a Dirac or a Majorana particle ? Does it have a magnetic moment ? \cite{minerva}.

The interest in neutrinos goes beyond the study of the particle's intrinsic properties, and extends to a variety of topics in astro-, nuclear and hadronic physics. Typical astrophysical examples include the understanding of the energy production in our sun, neutrino nucleosynthesis and the synthesis of heavy elements during the r-process, the influence of neutrinos on the dynamics of a core-collapse supernova explosion and the cooling of a proto-neutronstar \cite{bahcall,langanke}. In many astrophysical situations the neutrinos serve as messengers probing the interior of dense and opaque objects that otherwise remain inaccessible.
The influence of neutrinos even extends to cosmological questions such as  the role of neutrinos in the matter-antimatter asymmetry in the universe. In hadronic and nuclear physics, neutrino scattering can shed light on a lot of issues, including investigations of electroweak form factors, the study of the strange quark content of the nucleon and $\nu$-induced pion production \cite{minerva,boone,nutev}.

Despite the richness of phenomena they are involved in, neutrinos remain elusive particles, only weakly interacting and eager to escape detectors on the watch.
The presence of neutrinos, being chargeless particles, can only be inferred by  detecting  the secondary particles they create when colliding and interacting with matter. Nuclei are often used as neutrino detectors, providing relatively large cross sections that offer a broad variety of information.  
 As a consequence, a reliable interpretation of data involving neutrinos heavily counts on a detailed knowledge of the magnitude of neutrino-nucleus interactions under various circumstances. A precise knowledge of the energy and mass number dependence of the neutrino-nucleus cross section is essential to current and future measurements. The energies that neutrinos can transfer to nuclei depend on their origin. The 'low' energy regime extends to a few tens of MeV and relates to reactor, solar and supernova neutrinos. Atmospheric and accelerator neutrinos can carry energies from a few hundred MeV to several GeV's.

At intermediate energies (here defined as energies beyond the nuclear resonance region), neutrino-nucleus interactions have been studied within several approaches, investigating a variety of effects. The relativistic Fermi gas (RFG) model was employed in Refs.~\cite{horo,umino} to study the possibility of measuring strange-quark contributions to the nucleon form factors. 
The RFG takes into account the Fermi motion of the nucleons inside the nucleus, Pauli blocking and relativistic kinematics, but neglects several other effects. Ref.~\cite{polarlet} used a plane-wave impulse approximation description of the nuclear system to estimate polarization asymmetry effects in neutrino-induced nucleon knockout.
Relativistic nuclear effects were included in the calculations of Refs.~\cite{barbar,albe1,albe2,mai,meuccineut,meuccichar,vanderventel04}, using a relativistic shell model approach for the study of neutral-current and/or charged-current neutrino-nucleus scattering. In particular, in Refs.~\cite{albe1,albe2,mai} results in the relativistic plane-wave impulse approximation (RPWIA) were compared to RFG calculations. It is  shown that binding-energy effects tend to vanish as the energy increases. Going one step further in the complexity of the model calculation, the implementation of the final-state interactions (FSI) of the ejected nucleon has been achieved in different manners. In Ref.~\cite{bleve01} a phenomenological convolution model was applied to the RFG, showing that nucleon re-scattering can produce a reduction of the quasi-elastic cross section as large as $15\%$ at incoming neutrino energies of about 1 GeV. A description of FSI mechanisms through the inclusion of relativistic optical potentials is presented in Refs.~\cite{albe1,albe2,mai,meuccineut,meuccichar}. More specifically, Ref.~\cite{mai} studies the uncertainties derived from the use of different prescriptions for the potentials. A reduction of the cross section of at least $14\%$ is found at incoming neutrino energies of 1 GeV. In Refs.~\cite{meuccineut,meuccichar}, important FSI effects arise from the use of relativistic optical potentials within a relativistic Green's function approach. Apart from relativistic dynamics and FSI, other effects may have an impact on neutrino-nucleus reactions. In Refs.~\cite{kim1,kim2} the influence of relativistic nuclear structure effects, delta- and pion degrees-of-freedom, and RPA-type correlations on neutrino-scattering cross sections was examined.
Ref.~\cite{nieves} includes long-range correlations, FSI and Coulomb corrections in $^{12}C(\nu_{\mu},\mu^-)^{12}C^*$ calculations. An alternative method was proposed in Ref.~\cite{amaro}, where it was shown that a superscaling analysis of few-GeV inclusive electron scattering data allows one to predict charged-current neutrino cross sections in the nuclear resonance region, thereby effectively including delta isobar degrees-of-freedom.

In this paper we compute the single-nucleon knockout (often referred to as quasi-elastic (QE)) contribution to the inclusive neutrino-nucleus cross sections, for energies and nuclei relevant to proposals like Miner$\nu$a~\cite{minerva}, Miniboone~\cite{boone} and Finesse~\cite{finesse}. We consider that the large variety of relevant neutrino energies  and the tendency to study neutrino-nucleus interactions at increasing energies, necessitate the use of relativity. We employ two relativistic models for describing neutrino-nucleus scattering within the impulse approximation: the relativistic distorted-wave impulse approximation (RDWIA) developed by the Madrid-Seville group, and the relativistic multiple scattering Glauber approximation (RMSGA) developed by the Ghent group. Initially designed for the description of exclusive electron-nucleus scattering processes, both models have been succesfully tested against $A(e,e'p)$ data~\cite{Udi93,Udi96,Udi99,Udi01,cris04,lava04,lava05}. In addition, the nuclear transparencies predicted by these models have proven to be mutually consistent in the intermediate kinematic regime between 0.5 and 1 GeV nucleon kinetic energies where both of them are deemed reliable~\cite{lava04}. The RDWIA model used here has already been employed in several neutrino-nucleus calculations \cite{albe1,albe2,mai}. To our knowledge, this paper is the first report of a relativistic Glauber-inspired approach to neutrino-nucleus reactions. The aim of this work is twofold. First, the relativistic models available to date  predict different results in the limit of vanishing FSI, motivating a 'new round' of calculations. We investigate the plane-wave limit of the RDWIA and RMSGA approximations, aiming at providing benchmark RPWIA results. Second, we compute the effects of FSI within our models, paying special attention to the comparison between RDWIA and RMSGA results. It is well known that the inclusion of FSI within inclusive calculations requires a considerable computing effort. We propose a way to estimate FSI effects for the QE contribution to the inclusive neutrino-nucleus cross section using benchmark RPWIA results and transparency data obtained from $A(e,e'p)$ experiments. For the time being, the effects  of many-body currents, nucleon-nucleon correlations, and contributions beyond quasi-elastic scattering processes as multi-nucleon processes and pion production are neglected.

The outline of this paper is as follows. In Sec. \ref{sec:formalism} we present the RMSGA and RDWIA formalisms for the description of the neutral- and charged-current neutrino-nucleus scattering processes. Results of the numerical calculations are shown in Sec. \ref{sec:results}. Sec. \ref{sec:conclusions} summarizes our findings.

\begin{figure}[t]
  \begin{center} 	
    \includegraphics[width=10cm]{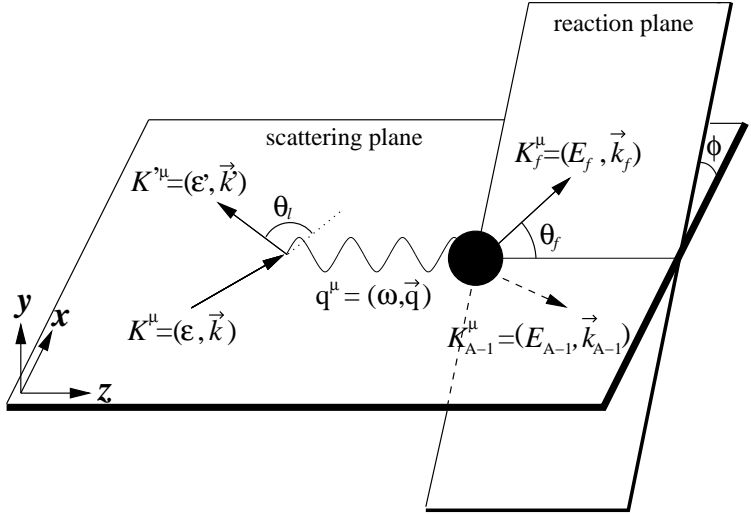}
   \caption{Kinematics for the quasi-elastic neutrino-nucleus scattering process}
\label{fig:kinema}
  \end{center} 
\end{figure}

\section{Formalism}
\label{sec:formalism}

We derive expressions for neutrino and antineutrino neutral-current (NC) reactions from nuclei which result in one emitted nucleon
\begin{equation}
\nu(\overline{\nu})+A\Longrightarrow\nu(\overline{\nu})+N+(A-1) .
\label{eq:NC}
\end{equation}
We also consider their charged-current (CC) counterparts
\begin{equation}
\nu(\overline{\nu})+A\Longrightarrow l(\overline{l})+N+(A-1) .
\label{eq:CC}
\end{equation}
Here, $l$
labels the flavor of the lepton, and $A$ represents a nucleus with mass number $A$.
The connection between electromagnetic and weak interactions makes that the analytical derivations follow the same lines as those used in electron-nucleus scattering. 
The main differences between neutrino and electron interactions stem from  the
intrinsic polarization of the neutrino due to the parity-violating character of the
weak interaction.  Moreover in weak interactions the focus is on inclusive processes, whereas exclusive processes play a predominant role in current subatomic research with electrons.


We describe these processes at lowest order in the electroweak interaction, i.e.  considering the exchange of one charged vector boson. Fig. \ref{fig:kinema} defines our conventions for the kinematical variables.
The
four-momenta of the incident neutrino and scattered lepton are labeled
 $K^{\mu}$ and $K'^{\mu}$.  Further,
$K^{\mu}_A$, $K^{\mu}_{A-1}$ and
$K^{\mu}_f$ represent the four-momenta of the target
nucleus, the residual nucleus and the ejected nucleon. The $xyz$
coordinate system is chosen such that the $z-$axis lies along the
momentum transfer $\vec{q}$, the $y-$axis along
$\vec{k}\times\vec{k'}$ and the $x-$axis lies in the scattering
plane. The hadron reaction plane is then defined by $\vec{k}_f$ and
$\vec{q}$. We adopt the standard convention $Q^2 \equiv
-q_{\mu}q^{\mu}$ for the four-momentum transfer.
\subsection{Quasi-elastic neutrino-nucleus cross section}
In the laboratory frame, the exclusive differential cross section for the processes specified in Eqs. (\ref{eq:NC}) and (\ref{eq:CC}) can be written as \cite{bjorken64}
\[ d\sigma = 
\frac{1}{\beta}\overline{\sum_{if}}|M_{fi}|^2
\frac{M_{l}}{\varepsilon'}\frac{M_{A-1}}{E_{A-1}}\frac{M_{N}}{E_{f}}
 d^3\vec{k}_{A-1} d^3\vec{k'}d^3\vec{k}_{f} \]
\begin{equation}
 \times
(2\pi)^{-5}\delta^4(K^{\mu}+K^{\mu}_A -
K'^{\mu}-K^{\mu}_{A-1}-K^{\mu}_f), 
\end{equation}
where $\overline{\sum_{if}}$ indicates sum and/or average over initial and final spins. Dealing with neutrinos, the relative initial velocity $\beta$ can trivially be put to 1. The factor $\frac{M_l}{\varepsilon'}$ stems from the normalization of the outgoing lepton spinor and becomes 1 for NC reactions.
Integrating over the unobserved momentum of the recoiling nucleus $\vec{k}_{A-1}$, as well as over $|\vec{k_f}|$, results in the following fivefold differential cross section for the $A(\nu,\nu'N)$, $A(\overline{\nu},\overline{\nu}'N)$, $A(\nu,lN)$ and $A(\overline{\nu},\overline{l}N)$ reactions 
\begin{equation}
\frac{d^5\sigma}{d\varepsilon'd^2\Omega_{l}d^2{\Omega_f}} =
\frac{M_{l}M_NM_{A-1}}{(2\pi)^5M_A\varepsilon'}k'^2k_f
f_{rec}^{-1}
\overline{\sum_{if}}|M_{fi}|^2,
\end{equation}
where $\Omega_{l}$ and ${\Omega_f}$ define the scattering direction of the outgoing lepton and the outgoing nucleon. The recoil factor $f_{rec}$ is given by 
\begin{equation}
f_{rec} = \frac{E_{A-1}}{M_A}\left|1 + \frac{E_f}{E_{A-1}}[1 - \frac{\vec{q}\cdot\vec{k}_f}{k_f^2}]\right|.
\end{equation}
The squared invariant matrix element $M_{fi}$ can be written as
\begin{equation}
\overline{\sum_{if}}|M_{fi}|^2 = \frac{G_F^2}{2}\left[\frac{M_B^2}{Q^2 +
M_B^2}\right]^2 l_{\alpha\beta}W^{\alpha\beta}.
\label{eq:mfi}
\end{equation}
Here, $M_B$ represents the mass of the $Z$-boson for NC reactions and that of the $W$-boson for CC processes. $G_F$ is the Fermi constant. For CC reactions the latter  has to
be multiplied with a factor $\cos{\theta_c}$, with $\theta_c$ the
Cabbibo angle, determining the mixing of the strong down and
strange quarks into the weak d-quark.
In the above expression the lepton tensor is defined as
\begin{equation}
l_{\alpha\beta} \equiv
\overline{\sum_{s,s'}}[\overline{u}_{l}\gamma_{\alpha}(1-\gamma_5)u_{l}]^{\dagger}
[\overline{u}_{\nu}\gamma_{\beta}(1-\gamma_5)u_{\nu}],
\end{equation} with $s$ and $s'$ the initial and final lepton spins.
The hadron tensor is given by
\begin{equation}
\label{eq:nucleartensor}
W^{\alpha\beta} = \overline{\sum_{if}}\langle\Delta^{\alpha
\mu}J_{\mu}\rangle^{\dagger}\langle\Delta^{\beta \nu}J_{\nu}\rangle=\overline{\sum_{if}}
\langle\cal{J}^{\alpha}\rangle^{\dagger}\langle\cal{J}^{\beta}\rangle,
\end{equation}
with the boson propagator
\begin{equation}
\Delta^{\mu\nu}= g^{\mu\nu} - \frac{q^{\mu}q^{\nu}}{M_B^2}.\label{delt}
\end{equation}
The quantity $\langle\cal{J}^{\alpha}\rangle$ in 
Eq.~(\ref{eq:nucleartensor}) can be written as
\begin{widetext}
\begin{equation}
\label{eq:current}
\left< \cal{J}^{\alpha} \right> \equiv
\left< (A-1)(J_R M_R),
K_f(E_f,\vec{k}_f)m_s\left|\Delta_{\alpha\mu}\hat{J}^{\mu} \right|A(0^+,g.s.)\right> ,
\end{equation}
\end{widetext}
with $\hat{J}^{\mu}$ the weak current
operator, $\left|A(0^+,g.s.)\right>$ the ground state of the target
even-even nucleus and $\left|(A-1)(J_R M_R)\right>$ the state
in which the residual nucleus is left. 
At the energies considered here, Eq.~(\ref{delt}) can approximately be written as $\Delta^{\mu\nu}\approx g^{\mu\nu}$, and the quantity $\langle \cal{J}^{\alpha}\rangle \approx \langle $$J^{\alpha} \rangle$, establishing the connection between the four-vector $\cal{J}^{\alpha}$ and the nuclear current operator.
In the extreme
relativistic limit, the contraction
of the lepton tensor $l_{\alpha\beta}$ with the nuclear one
$W^{\alpha\beta}$ in Eq.~(\ref{eq:mfi}) can be cast in the form \cite{donnelly86}~:
\begin{eqnarray}
\frac{d^5\sigma}{d\varepsilon'd^2\Omega_{l}d^2{\Omega_f}} &=& \frac{M_NM_{A-1}}{(2\pi)^3M_A}k_ff_{rec}^{-1} \sigma_M^{Z,\,W^{\pm}} \nonumber\\&&\hspace*{-3cm}\times
\left
[v_LR_L + v_TR_T + v_{TT}R_{TT}\cos{2\phi}\right.\nonumber\\&&\hspace*{-2.5cm}\left.+ v_{TL}R_{TL}\cos{\phi} + h(v_T'R_T' + v_{TL}'R_{TL}'\cos{\phi})\right],
\label{eq:cross}
\end{eqnarray}
with $\sigma_M$ defined by
\begin{equation}
\sigma_M^Z =
  \left(\frac{G_F\cos(\theta_l/2)\varepsilon'M_Z^2}{\sqrt{2}\pi(Q^2+M_Z^2)}\right)^2, 
\end{equation}
for NC reactions and 
\begin{equation}
\sigma_M^{W^\pm} =\sqrt{1 -\frac{M_{l}^2}{\varepsilon'^2}}
  \left(\frac{G_F\cos(\theta_c)\varepsilon'M_W^2}{2\pi(Q^2 +
    M_W^2)}\right)^2 ,
\end{equation}
for CC reactions.
In these equations, $\theta_l$ is
the angle between the direction of the incident and the scattered lepton's momentum and $\phi$ the
azimuthal angle of the reaction plane (see Fig. \ref{fig:kinema}). In Eq.~(\ref{eq:cross}), $h = -1$ ($h = +1$) corresponds to the helicity of the incident neutrino (antineutrino). For NC
reactions, the lepton kinematics is contained in the kinematic
factors
\begin{eqnarray}
v_L &=& 1,
\\
v_T &=& \tan^2{\frac{\theta_l}{2}} + \frac{Q^2}{2|\vec{q}|^2},
\\
v_{TT} &=& - \frac{Q^2}{2|\vec{q}|^2},
\\
v_{TL} &=& -\frac{1}{\sqrt{2}}\sqrt{\tan^2{\frac{\theta_l}{2}} + \frac{Q^2}{|\vec{q}|^2}},
\\
v_T' &=& \tan{\frac{\theta_l}{2}}\sqrt{\tan^2{\frac{\theta_l}{2}} + \frac{Q^2}{|\vec{q}|^2}},
\\
v_{TL}' &=& \frac{1}{\sqrt{2}}\tan{\frac{\theta_l}{2}}.
\end{eqnarray}
The corresponding response functions read 
\begin{widetext}
\begin{eqnarray}
R_L &=& \left|\langle {\cal J}^0(\vec{q})\rangle - \frac{\omega}{|\vec{q}|}\langle
{\cal J}^z(\vec{q})\rangle\right|^2,\label{rl}
\\
R_T &=& \left|\langle {\cal J}^+(\vec{q})\rangle\right|^2 + \left|\langle {\cal J}^{-}(\vec{q})\rangle\right|^2,
\\
R_{TT}\cos{2\phi} &=& 2\Re\left\{\langle {\cal J}^+(\vec{q})\rangle^*\langle
{\cal J}^-(\vec{q})\rangle\right\},
\\
R_{TL}\cos{\phi} &=& -2\Re\left\{\left[\langle {\cal J}^0(\vec{q})\rangle -
  \frac{\omega}{|\vec{q}|}\langle {\cal J}^0(\vec{q})\rangle\right]\left[\langle
  {\cal J}^+(\vec{q}) \rangle - \langle {\cal J}^-(\vec{q})\rangle\right]^* \right\},
\\
R_T' &=& \left|\langle {\cal J}^+(\vec{q})\rangle\right|^2 - \left|\langle {\cal J}^-(\vec{q})\rangle\right|^2,
\\
R_{TL}'\cos{\phi} &=& -2\Re\left\{\left[\langle {\cal J}^0(\vec{q})\rangle -
  \frac{\omega}{|\vec{q}|}\langle {\cal J}^z(\vec{q})\rangle\right]\left[\langle
  {\cal J}^+(\vec{q})\rangle + \langle {\cal J}^-(\vec{q})\rangle\right]^* \right\},\label{rtl}
\end{eqnarray}
\end{widetext}
where  $\langle \vec{\cal J}(\vec{q})\rangle$ is expanded in terms of unit spherical vectors $\vec{e}_m$
\begin{equation}
\vec{e}_0 = \vec{e}_z, ~~~~~~ \vec{e}_{\pm 1} = \mp
\frac{1}{\sqrt{2}}(\vec{e}_x \pm i\vec{e}_y).
\end{equation}

For CC reactions, the mass of the outgoing lepton has to be taken into account. This results in the following substitutions (see also for instance \cite{umino}) 
\begin{widetext}
\begin{eqnarray}
v_T &=& 1 -\sqrt{1 - \frac{M_{l}^2}{\varepsilon'^2}}\cos{\theta_l} +
  \frac{\varepsilon\varepsilon'}{|\vec{q}|^2}\left(1 - \frac{M_{l}^2}{\varepsilon'^2}\right)
  \sin^2{\theta_l},
\\
v_{TT} &=& - \frac{\varepsilon\varepsilon'}{|\vec{q}|^2}\left(1 - \frac{M_{l}^2}{\varepsilon'^2}\right)\sin^2{\theta_l},
\\
v_{TL} &=&
\frac{\sin{\theta_l}}{\sqrt{2}|\vec{q}|}(\varepsilon +
\varepsilon'),
\\
v_T' &=&  {\frac{\varepsilon + \varepsilon'}{|\vec{q}|}}\left(1 -\sqrt{1 -
  \frac{M_{l}^2}{\varepsilon'^2}}\cos{\theta_l}\right)-\frac{M_{l}^2}{\varepsilon'|\vec{q}|},
\\
v_{TL}'&=& -\frac{\sin{\theta_l}}{\sqrt{2}}\sqrt{1 -
\frac{M_{l}^2}{\varepsilon'^2}}.
\end{eqnarray}
Furthermore
\begin{equation}
R_{TL}\cos\phi=2\Re\biggl\{\left[\langle
  {\cal J}^0(\vec{q})\rangle-\frac{\omega+M_{l}^2}{|\vec{q}|}\langle
  {\cal J}^z(\vec{q})\rangle\right]\left[\langle {\cal J}^+(\vec{q})\rangle 
-\langle {\cal J}^-(\vec{q})\rangle\right]^* \biggr\},
\end{equation}
and
\begin{equation}
v_LR_L = v_L^0R_L^0 + v_L^zR_L^z + v_L^{0z}R_L^{0z},
\end{equation}
with
\begin{equation}
R_L^0=\left|\langle {\cal J}^0(\vec{q})\rangle\right|^2, \;\;\;R_L^z=\left|\langle {\cal J}^z(\vec{q})\rangle\right|^2,\;\;\;R_L^{0z}=-2\Re\left\{\langle {\cal J}^0(\vec{q})\rangle\langle {\cal J}^z(\vec{q})\rangle^*\right\},
\end{equation}
and
\begin{eqnarray}
v_L^0&=&
\left[1 + \sqrt{1 -
  \frac{M_{l}^2}{\varepsilon'^2}}\cos{\theta_l}\right], 
  \\ v_L^z&=&  \left[ 1 +\sqrt{1 -
  \frac{M_{l}^2}{\varepsilon'^2}}\cos{\theta_l} -
  \frac{2\varepsilon\varepsilon'}{|\vec{q}|^2}\left(1 -
  \frac{M_{l}^2}{\varepsilon'^2}\right)
  \sin^2{\theta_l}\right],
   \\  v_L^{0z}&=&   \left[
  \frac{\omega}{|\vec{q}|}\left(1+\sqrt{1 -
  \frac{M_{l}^2}{\varepsilon'^2}}\cos{\theta_l}\right) +
  \frac{M_{l}^2}{\varepsilon'|\vec{q}|}\right].
\end{eqnarray}
\end{widetext}
The expressions for $R_T$, $R_{TT}$, $R'_{T}$ and $R'_{TL}$ remain unaltered.

So far, a precise knowledge of the kinematic
variables at the lepton vertex was assumed. In practice, this information is not attainable in typical neutrino
scattering experiments. Indeed, in NC reactions, the scattered
lepton is chargeless and remains undetected. In CC processes, on the other hand, detection of the final lepton is possible and its
energy and momentum could in principle be measured. However, due to limited control on the incoming neutrino energies, the energy-momentum balance at the lepton vertex cannot be precisely determined. In order to get the QE neutrino-nucleus cross section, we integrate over the phase space of the scattered lepton ($d^2\Omega_l$) and the outgoing nucleon ($d^2\Omega_f(\theta_f,\,\phi)$). For
the latter, integration over the azimuthal angle $\phi$ yields a factor 
$2\pi$, whilst only the $\phi$-independent terms survive due to
symmetry properties. 
This yields
\begin{widetext}
\begin{equation} 
\label{eq:intcross}
\frac{d\sigma}{d\varepsilon'} = \frac{M_NM_{A-1}}{(2\pi)^3M_A}4\pi^2 
 \int\sin{\theta_l}d\theta_l\int\sin{\theta_f}d\theta_f k_ff_{rec}^{-1} \sigma_M[v_LR_L + v_TR_T + hv_T'R_T'] \; .
\end{equation} 
\end{widetext}
In practice, we compute the response functions for all single-particle levels in the target nucleus, and obtain $d\sigma/d\varepsilon'$ by summing over all these.
\subsection{Nuclear current}
Obviously, the determination of the response functions requires knowledge of the nuclear current matrix elements (\ref{eq:current}). We describe the neutrino-nucleus nucleon-knockout reaction within the impulse approximation, assuming that the incident neutrino interacts with only one nucleon, which is subsequently emitted. The nuclear current is written as a sum of single-nucleon currents. The wave functions for the target and the residual nuclei are described in terms of an independent-particle model. Then, the transition matrix elements can be cast in the following form :
\begin{equation}
\label{eq:relcurrent}
  \langle J^{\mu}
  \rangle = \int
  d\vec{r} \; \overline{\phi}_F(\vec{r})\hat{J}^{\mu}(\vec{r})e^{i\vec{q}.\vec{
      r}}\phi_{B}(\vec{r}) \; ,
 \end{equation}
where $\phi_{B}$ and $\phi_F$ are relativistic bound-state and
scattering wave functions. Further, $\hat{J}^{\mu}$ is the
relativistic one-body current operator modeling the coupling between
the virtual $Z^0$ or $W^{\pm}$ boson and a bound nucleon. The
relativistic bound-state wave functions are obtained within the
Hartree approximation to the $\sigma$-$\omega$ model \cite{serot86}.
The quantum-field theoretical problem is solved in the standard mean-field approximation    replacing the meson field operators by   their
expectation values. The resulting eigenvalue equations of the
relativistic mean-field theory can be solved exactly.  The
corresponding bound-state wave functions $\phi_{B}$ are four-spinors and
can  formally be written as
\begin{equation}
\phi_{B}(\vec{r})=\left( \begin{array}{@{\hspace{0pt}}c@{\hspace{0pt}}}
                \frac {i G_{n_{B} \kappa_{B}}(r)} {r} \mathcal{Y}
                _{\kappa_{B} m_{B}}
                 (\Omega_r,\vec{\sigma}) \\
                \frac {- F_{n_{B} \kappa_{B}}(r)} {r} 
              \mathcal{Y} _{- \kappa_{B} m_{B}}(\Omega_r,\vec{\sigma}) \\
             \end{array} \right) \, ,
\label{bwf}
\end{equation}
with $\mathcal{Y} _ {\kappa _{B} m_{B}}(\Omega_r, \vec{\sigma})$ the
familiar spin spherical harmonics. 

We  use a
relativistic one-body vertex function of the form: 
\begin{widetext}
\begin{equation}
  J^{\mu} = F_1(Q^2)\gamma^{\mu} +
i\frac{\kappa}{2M_N}F_2(Q^2)\sigma^{\mu\nu}q_{\nu} +
G_A(Q^2)\gamma^{\mu}\gamma_5 + \frac{1}{2M_N} G_P(Q^2)q^{\mu}\gamma_5,
\label{eq:nuclearoperator}
\end{equation}
\end{widetext}
with $\kappa$ the anomalous magnetic moment.  
The weak vector form factors $F_1$ and $F_2$ can be related to the
corresponding electromagnetic ones for protons
$(F_{i,p}^{EM})$ and neutrons $(F_{i,n}^{EM})$ by the conserved vector
current (CVC) hypothesis. For proton knockout they are given by
\begin{widetext}
\begin{equation}
\label{eq:dirac}
F_i = 
\begin{cases}
\left(\frac{1}{2}-2\sin^2{\theta_W}\right)F_{i,p}^{EM} - \frac{1}{2} F_{i,n}^{EM}  &
\text{for NC reactions}, \\
(F_{i,p}^{EM} - F_{i,n}^{EM}) & \text{for CC reactions}, 
\end{cases}
\end{equation}
\end{widetext}
with  $\theta_W$ the
Weinberg angle defined by $\sin^2{\theta_W} = 0.2224$. For neutron knockout the weak vector form factors result from the exchange of the subindexes $p$ and $n$ in Eq.~(\ref{eq:dirac}). A standard dipole parametrization is adopted for the vector form factors. 
 
The axial form factor for proton knockout is expressed as
\begin{equation}
G_A =
\begin{cases}
  - \frac{g_A}{2}G & \text{for NC reactions},
    \\ {g_A} G& \text{for CC
    reactions}
\end{cases}
\label{eq:ga}
\end{equation}
where $g_A \mbox{=} 1.262$, and $G=(1+Q^2/M^2)^{-2}$ with $M=1.032$ GeV. For neutron knockout a minus sign must be added to Eq. (\ref{eq:ga}).  

 
The Goldberger-Treiman relation allows one to write the pseudoscalar
form factor as
\begin{equation}
G_P(Q^2) = \frac{2M_N}{Q^2 + m_{\pi}^2}G_A(Q^2),
\end{equation}
where $m_{\pi}$ denotes the pion mass.
The contribution of this form factor, being proportional to the
mass of the scattered lepton, vanishes for NC reactions. 
\subsection{Final-state interactions in relativistic models}

We now turn to the question of computing a relativistic scattering
wave function for the outgoing nucleon. Including nucleon-nucleus FSI is a long-standing issue in
theoretical $A(e,e'p)$ investigations. For kinetic energies up
to around $1$ GeV, most calculations have traditionally been performed within a
so-called distorted-wave impulse approximation model (DWIA), where the final nucleon scattering state is computed with
the aid of proton-nucleus optical potentials. For proton kinetic energies
 above $1$ GeV parameterizations of
these potentials within the context of Dirac phenomenology are not readily at hand. Furthermore, beyond this energy the use of optical potentials for modeling FSI processes does not seem very natural in view of the highly inelastic and diffractive properties of the underlying nucleon-nucleon scattering process. In this energy regime, the Glauber model, which is
a multiple-scattering extension of the eikonal approximation, offers a
valid and economical alternative for describing FSI \cite{glauber70}. 
In a Glauber
framework, the FSI effects are computed directly  from the elementary
nucleon-nucleon scattering data. Below, we give a brief outline of the main features of both models.

Within the RDWIA framework \cite{Udi93,Udi96,Udi99,Udi01}, $\phi_F$ in Eq. (\ref{eq:relcurrent}) is a scattering solution to a
Dirac-like equation, which includes scalar and vector complex optical potentials obtained by fitting elastic $pA$ 
scattering data. The real part of these potentials describes the rescatterings of the ejected nucleon. The imaginary part accounts for the absorption into unobserved channels. The scattering wave function, expressed in terms of
a partial-wave expansion in configuration space, reads
\begin{eqnarray}
\phi_F({\vec{r}})=4\pi\sqrt{\frac{E_f+M_N}{2E_f}}\sum_{\kappa \mu
m}e^{-i \delta^\ast_\kappa} i^\ell\langle\ell m \frac{1}{2} s_f|j \mu
\rangle\nonumber \\ \times Y_\ell^{m\ast}(\Omega_{k_f})\Psi_\kappa^{\mu}({\vec{r}}) \, ,
\label{dwf}
\end{eqnarray}
where $\Psi_\kappa^{\mu}({\vec{r}})$ are four-spinors of the same form
as in Eq.~(\ref{bwf}). The phase-shifts $\delta^\ast_\kappa$ and radial functions are complex because of the complex potentials. The outgoing nucleon spin is denoted as $s_f$. In this work we use the relativistic global optical potential corresponding to the energy and target mass-dependent parametrization (EDAD1) of Ref. \cite{Cooper93}.

The Glauber approach relies on the eikonal and the frozen-spectators
approximation.  It allows  to formulate a full-fledged
multiple-scattering theory for the emission of a ``fast'' nucleon from
a composite system consisting of $A-1$ temporarily ``frozen'' nucleons.  
Ref.~\cite{ryckebusch03} provides a detailed outline of  a relativistic and unfactorized formulation
of Glauber multiple scattering theory.  In this
approach, coined RMSGA, the scattering wave function in the matrix
element of Eq.~(\ref{eq:relcurrent}) adopts  the form
\begin{equation}
\label{eq.:transition}
\phi_F({\vec{r}}) \equiv  \mathcal{G}
(\vec{b},z)\phi _{k_f, \; s_f}(\vec{r}) \,
\end{equation}
 where $\phi_{k_f, \; s_f}$ is a relativistic plane wave.  The impact
 of the FSI mechanisms on the scattering wave function is contained in
 the scalar Dirac-Glauber phase $\mathcal{G}(\vec{b},z)$
\begin{equation}
\label{eq.:glauberphase}
\mathcal{G}(\vec{b},z)= \prod_{\alpha \neq B} \biggl[ 1-\int
d\vec{r}~'|\phi_{\alpha}(\vec{r}~')|^2 \theta(z'-z)\Gamma(\vec{b}'
-\vec{b}) \biggr] ,
\end{equation}
where the product over $\alpha (n, \kappa, m)$ extends over all
occupied single-particle states in the target nucleus, excluding
the one from which the nucleon is ejected.  The profile function for
$NN$ scattering is defined in the standard manner
\begin{equation}
\label{eq.:profile}
\Gamma(\vec{b})=
\frac {\sigma_{NN}^{tot} (1-i\epsilon_{NN})} 
{4\pi\beta^2_{NN}} \exp(\frac{-b^2}{2\beta^2_{NN}})\;.
\end{equation}
The parameters $\sigma_{NN}^{tot}$, $\beta_{NN}$ and $\epsilon_{NN}$
depend on the ejectile energy. Values for the parameters fitted to the $pp$ and $pn$ data can be
found in Ref.~\cite{pdg}. The neutron-neutron scattering parameters are assumed identical to the proton-proton ones.

As the
integrations in Eq.~(\ref{eq:intcross}) would require an enormous numerical
effort, we introduce an additional averaging over the positions of the
spectator nucleons. This procedure amounts to replacing in
Eq.~(\ref{eq.:glauberphase}) the characteristic spatial
distributions of each of the spectator nucleons by an average density
distribution for the target nucleus
\begin{eqnarray}
\label{eq:glauberthick}
\mathcal{G}(\vec{b},z) &\approx&   
\Biggl\{1 - 
\frac{\sigma^{tot}_{NN}( 1 - i \epsilon _{NN})} {4\pi \beta _{NN}^2}\nonumber \\ &\times& 
\int _{0} ^{\infty} b'db' T_B(b',z) \exp \left[ -\frac{(b - b')^2 }{2
    \beta _{NN} ^2} \right] \nonumber \\ &\times&
\int _{0} ^{2 \pi} d \phi_{b'}
\exp \biggl[ \frac{-2b b'}{\beta_{NN}^2} {\sin}^2 \left( \frac {\phi_{b} -\phi_{
b'}} {2} \right) \biggr]
\Biggr\}^{A-1} \; .
\end{eqnarray} 
The function $T_B(b',z)$ which was introduced in the above expressions
is known as the ``thickness function'' and reads 
\begin{equation}
\label{eq:thick}
T_B(b',z) = \frac{1}{A}\int _{- \infty}
^{+ \infty} dz' \theta (z' -z) \rho_B(r'(b',z')),
\end{equation}
where the relativistic radial baryon density $\rho_B(r)$ is defined in
the standard fashion
\begin{eqnarray}
\rho_B(r) &\equiv& \langle \overline{\Psi_A^{gs}}\gamma_0
\Psi_A^{gs}\rangle = \sum_{\alpha}\int d\vec{\sigma}d\Omega
(\phi_{\alpha}(\vec{r},\vec{\sigma}))^{\dagger}(\phi_{\alpha}(\vec{r},\vec{\sigma}))
\nonumber \\
&=& \sum_{n\kappa}\frac{(2j +1)}{4\pi r^2}\biggl[\left|G_{n\kappa}(r)
  \right|^2 + \left|F_{n\kappa}(r)
  \right|^2 \biggr],
\end{eqnarray}
and the sum over $n\kappa$ extends over all occupied states. For exclusive $A(e,e',p)$ processes, where the quantum numbers of the residual nucleus are well defined, the thickness-function approximation of Eq.~(\ref{eq:glauberthick}) provided results which approach the exact ones obtained with the expression of Eq.~(\ref{eq.:glauberphase}) \cite{ryckebusch03}.  Here, we only deal with inclusive cross sections, obtained after incoherently summing over nucleon emission from all possible single-particle shells.  Under these circumstances, one can expect that the thickness-function approximation becomes an even better one.

The RDWIA and RMSGA codes were developed independently and adopt distinctive numerical techniques to compute the scattering wave functions
and the corresponding matrix elements of
Eq.~(\ref{eq:relcurrent}). The RDWIA code employs a partial-wave
expansion to solve the Dirac equation for the ejectile. The cylindrical symmetry of the Glauber phase of Eq.~(\ref{eq.:glauberphase}) prohibits any meaningful use of this technique in the RMSGA calculations. Instead, the multi-dimensional integrals are computed numerically. In the limit of vanishing FSI mechanisms, i.e. within RPWIA, though, both models should
yield identical results. In the Glauber picture this limit is reached
by putting the Glauber phase of Eq.~(\ref{eq.:transition}) equal to unity. In the RDWIA picture, the effect of FSI can be made vanishing by nullifying the optical potentials. Then, the computed partial waves sum to a relativistic plane
wave. Convergence of the partial wave expansion was tested against the analytical plane-wave result.

The models described above were initially developed for the description of exclusive $A(e,e'p)$ processes, for which an excellent agreement between theoretical calculations and data has been achieved~\cite{Udi93,Udi96,Udi99,Udi01,cris04,lava05}. It is clear that inclusive neutrino scattering cross sections include
  contributions which fall beyond the scope of the RDWIA and RMSGA
  models. Both the RDWIA and RMSGA are confined to those processes
  where the scattering of a neutrino from a nucleus causes a single
  nucleon to escape, thereby exciting the residual nucleus in a state
  at missing energies below 80 MeV and a predominant single-hole
  nuclear structure with respect to the ground state of the target
  nucleus. We refer to such processes as "elastic" ones and wish to
  stress that they include proton and neutron knockout from
  the deepest lying $1s$ up to the Fermi level.  Inelastic single-nucleon
  knockout channels populating more complex states in the residual A-1
  nucleus are excluded from our calculations. So are multi-nucleon
  knockout channels and channels involving a pion.  In that sense, the
  RDWIA and RMSGA predictions for the inclusive neutrino-nucleus cross
  sections should be interpreted as a lower limit of the
  single-nucleon knockout contribution.

\section{Results}
\label{sec:results}

We present results for QE neutrino scattering from $^{12}$C and $^{56}$Fe, which are nuclei well suited for neutrino detection. The calculations span incident neutrino energies from 150 MeV up to 5 GeV. From about 200 MeV to 1 GeV, the quasi-elastic nucleon knockout is expected to be the dominant contribution to the neutrino-nucleus cross section. At higher energies, the relative contribution of the inelastic channels, mainly those involving an intermediate delta resonance and pion production, is expected to become increasingly dominant in the inclusive process ~\cite{ahrens,paschos}. Ref.~\cite{ahrens} indicates
that in the neutrino energy range from 0.7 to 5 GeV reaction channels
involving a pion contribute for $15 \%$ to the total cross section.

In order to make the comparisons between the RDWIA and RMSGA calculations as meaningful as possible, all the ingredients in the calculations not related to FSI, as
those concerning the implementation of relativistic dynamics and
nuclear recoil effects, are kept identical. In particular, both pictures adopt the W1 parametrization \cite{furnstahl97} for the
different field strenghts in determining the bound-state wave
functions. Accordingly, the RDWIA and RMSGA only differ in their assumptions regarding the treatment of FSI. 

It speaks for itself that before embarking on the study of effects like FSI, pion production, the role of the delta in the
medium, multi-nucleon knock-out, the strangeness content of the nucleon
..., it is absolutely essential to possess reliable baseline RPWIA
cross sections with a numerical accuracy of a few percent. To this
purpose, before turning to the study of the role of FSI mechanisms, we first
investigate the RPWIA limit of the RMSGA and RDWIA
models. These predictions will be
compared and confronted with other RPWIA results which
made their way to literature recently \cite{meuccineut,vanderventel04}.

\begin{figure*}[h]
  \begin{center} 	
    \includegraphics[width=12cm]{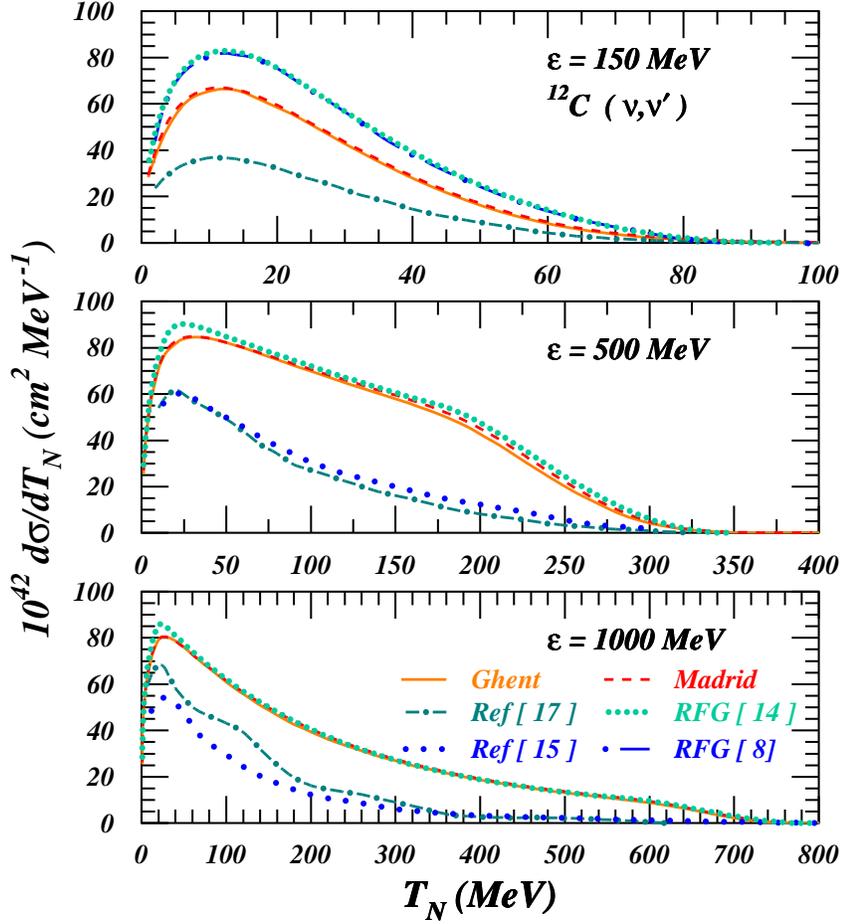}
    \caption{Neutral current $^{12}$C$({\nu},{\nu}')$ cross sections
   as a function of the outgoing nucleon kinetic energy $T_N$ at different incoming neutrino energies. The solid
   (dashed) lines represent the RPWIA results of the Ghent (Madrid)
   group. The short-dot-dashed lines show the RPWIA results of Ref.~\cite{vanderventel04}, and the long-dotted lines those of Ref.~\cite{meuccineut}. The short-dotted (long-dot-dashed) line shows the
   predictions of the  RFG model of Ref.~\cite{mai} (Ref.~\cite{horo})
   with a binding-energy correction of $27$ MeV.}
\label{fig.:c12crossncplane150}
  \end{center} 
\end{figure*}

\subsection{RPWIA}

Fig.~\ref{fig.:c12crossncplane150} shows the results of various RPWIA
calculations for $^{12}$C$(\nu,\nu')$ at $150$, $500$ and $1000$ MeV. We observe that the plane-wave limits of our RMSGA and RDWIA formalisms are in excellent agreement. The remaining differences, smaller than 2-3$\%$, can be attributed to the distinctive numerical techniques. This comparison lends us confidence about the consistency of the two types of calculations and the reliability of the adopted numerical techniques.

The fact that our models provide almost identical RPWIA results may seem trivial. As can be appreciated via Fig.~\ref{fig.:c12crossncplane150}, however, our RPWIA predictions disagree with the ones of Refs.~\cite{meuccineut} and \cite{vanderventel04}. Although the RPWIA calculations of Refs.~\cite{meuccineut} and \cite{vanderventel04} are mutually consistent at $\varepsilon=500$ MeV, this is no longer the case at $\varepsilon=1$ GeV. In the search for the origin of the discrepancies between our and other RPWIA calculations, differences in the nuclear current can be ruled out. The current operator of Eq.~(\ref{eq:nuclearoperator})  used along this work is formally identical to the one mentioned in Refs.~\cite{meuccineut} and
\cite{vanderventel04}, and the same holds for the form-factor parameterization. Only the bound-state wave functions used in Refs.~\cite{meuccineut} and
\cite{vanderventel04} differ from ours. We have performed cross section calculations with various parameterizations for the bound state wave functions, and found almost negligible differences. 

\begin{figure*}[h]
  \begin{center} 	
    \includegraphics[width=12cm]{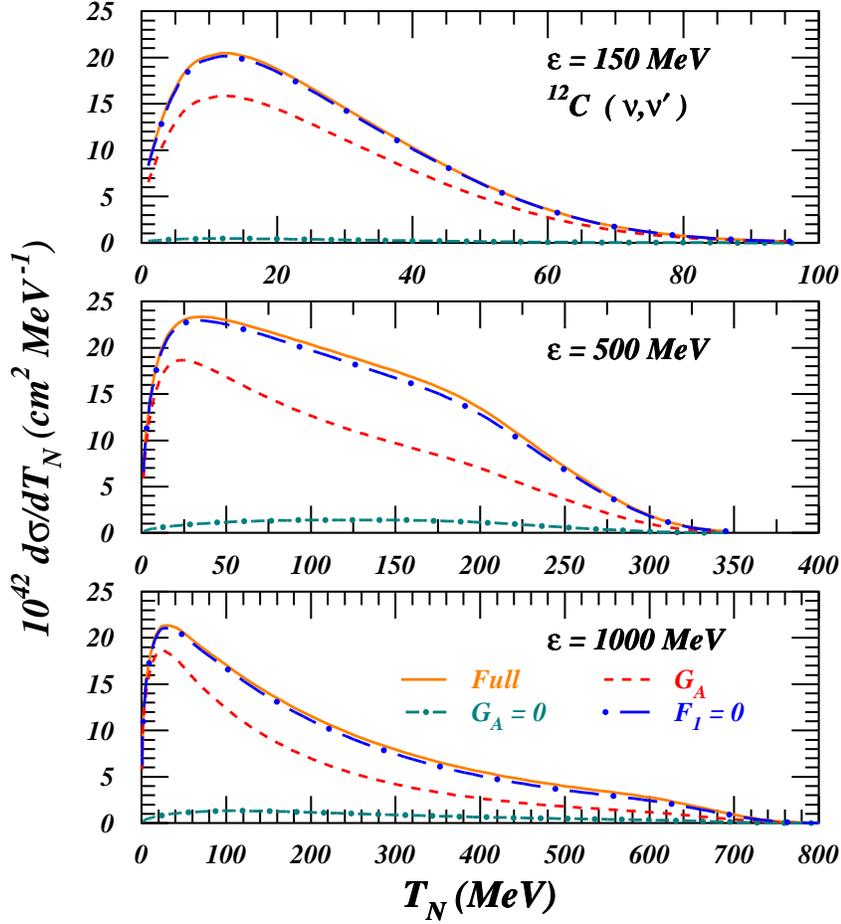}
    \caption{Effect of the different form factors on the neutral current $^{12}$C$({\nu},{\nu}')$ cross sections
   as a function of $T_N$ at different incoming neutrino energies. The solid lines represent the full RPWIA results. The long-dot-dashed (short-dot-dashed) lines show the results with $F_1=0$ ($G_A=0$). The dashed lines show the cross section when only $G_A$ is considered.}
   \label{fig.:c12formfactors}
  \end{center} 
\end{figure*}

The role of the various terms $F_1$, $F_2$ and $G_A$ in Eq.~(\ref{eq:nuclearoperator}) in the NC differential cross section was investigated in Ref.~\cite{vanderventel04}. The results were shown for proton knockout from the $1p_{3/2}$ orbital of $^{12}$C, at incident neutrino energies of 150, 500 and 1000 MeV. In Fig.~\ref{fig.:c12formfactors} we analyse the contribution of the $F_1$, $F_2$ and $G_A$ form factors in our cross sections under the same circumstances. First, let us observe that the calculations performed by setting $F_1=0$ (long-dot-dashed lines) almost reproduce the full cross sections (solid lines). This shows, in agreement with the outcomes of Ref.~\cite{vanderventel04}, that the contribution of the Dirac form factor is very small. The cross section can then be very well approximated as a sum of three terms: one proportional to ($G_A$)$^2$, a second proportional to ($F_2$)$^2$, and a third proportional to the interference of $G_A$ and $F_2$ contributions. The term proportional to ($G_A$)$^2$ (dashed lines) is very similar to the corresponding one in Fig.~11 of Ref.~\cite{vanderventel04}. The same holds for the cross sections obtained by nullifying $G_A$ (short-dot-dashed lines), whose behaviour is almost entirely given by $F_2^2$ term. Thus, the fact that our curves neglecting the $F_1$ contribution differ from those in Ref.~\cite{vanderventel04} can mainly be attributed to the $G_AF_2$ interference term. Furthermore, the results seem to differ in the sign of this term. As a matter of fact, changing the sign of the $G_AF_2$ term in our calculations, the cross sections follow closely the ones of Ref.~\cite{vanderventel04}. At $500$ and $1000$ MeV the differential cross sections of Ref.~\cite{vanderventel04} display some oscillations as a function of $T_N$. As can be appreciated from Fig.~\ref{fig.:c12crossncplane150} and \ref{fig.:c12formfactors}, we find no indications for these oscillations. Recently, the authors of Ref.~\cite{vanderventel04} have extended their work to calculate CC neutrino cross sections \cite{vanderventel05}. We remark that for this type of neutrino reactions the magnitude of the cross sections is in agreement with the RPWIA limit of the models presented here.

It is well known that binding-energy effects tend to vanish with increasing energies. Accordingly, a description of the $\nu$-nucleus scattering process in terms of a RFG model is expected to approach the RPWIA predictions at high incoming neutrino energies. This is observed in Fig.~\ref{fig.:c12crossncplane150}, when  comparing the RFG results of Refs.~\cite{horo,albe1,mai} with our RPWIA predictions. At $\varepsilon=150$ MeV, our RPWIA cross sections are approximately $15 \%$ smaller than the RFG ones. The RFG result from Refs.~\cite{albe1,mai} closely follows our RPWIA results at 500 MeV, the agreement at 1 GeV being remarkably good. The observed similarity between the independent RFG predictions of Refs.~\cite{horo,albe1,mai} and our RPWIA results lends us additional confidence that the RPWIA results presented here can serve as benchmark calculations.

\subsection{The effect of FSI : RMSGA and RDWIA approaches}

\begin{figure*}[t]
  \begin{center} 	
   \includegraphics[width=15cm]{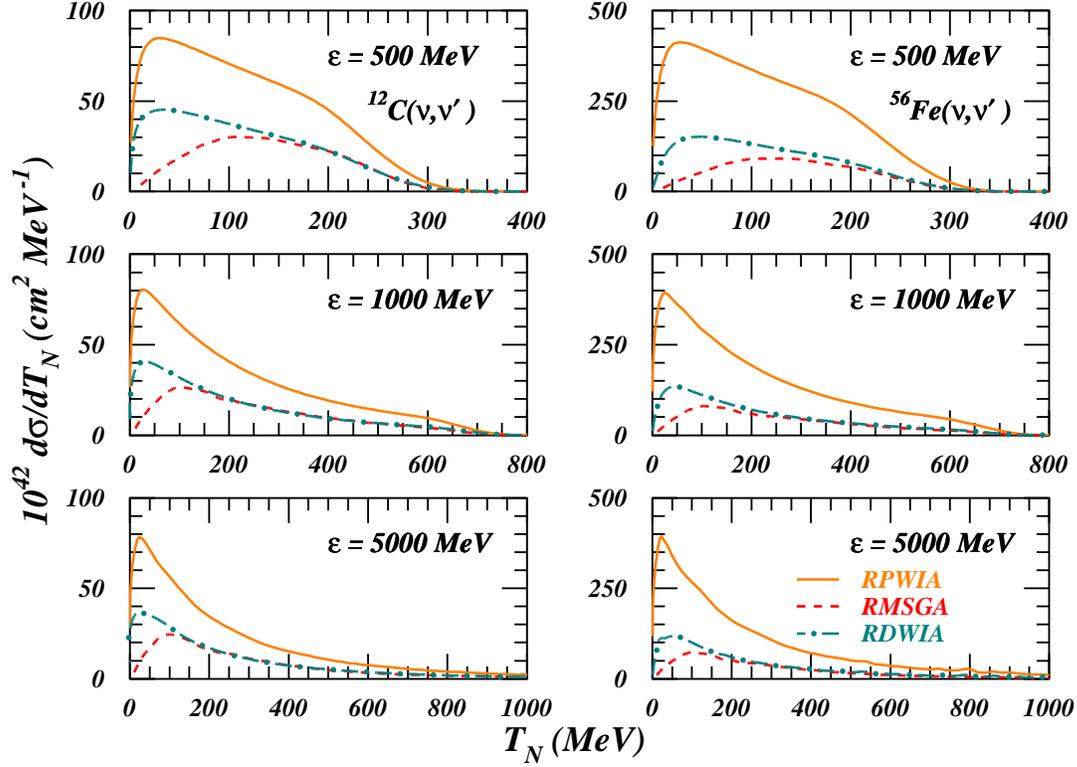}
   \caption{Neutral current $^{12}$C$({\nu},{\nu}')$ (left panels) and
   $^{56}$Fe$({\nu},{\nu}')$ (right panels) cross sections as a
   function of $T_{N}$ at different incoming neutrino energies. The solid
   lines represent the RPWIA predictions of the Madrid group, in agreement with those of the Ghent one. The dashed (dot-dashed) lines implement the effect of
   FSI within the RMSGA (RDWIA) framework. }
\label{fig.:c12nc}
  \end{center} 
\end{figure*}

Let us now turn our attention to the effect of FSI. NC $\nu$-nucleus cross sections obtained within RDWIA and RMSGA are displayed in Fig.~\ref{fig.:c12nc}. The calculations correspond to $^{12}$C and $^{56}$Fe targets, and incoming energies of $500$, $1000$,
and $5000$ MeV. Focusing on the results of the RDWIA model, the inclusion of the complex optical potential reduces the RPWIA results by
nearly $40-50 \%$ for $^{12}$C. As expected, the global effect of FSI increases with growing atomic
number, and reductions of over $60 \%$ are obtained for $^{56}$Fe. The presence of the
  imaginary term in the optical potential is likely to lead to an
  underestimation of the single-nucleon knockout contribution to the
  inclusive cross section.  Indeed, in inclusive measurements all
  possible final channels are included, whilst the RDWIA and RMSGA
  calculations are confined to "elastic" single-nucleon knockout.

Traditionally, Glauber-inspired models have been esteemed to provide reliable results at high energies, as they rely on the eikonal approximation. A very striking outcome of Fig.~\ref{fig.:c12nc} is that, for integrated quantities as the ones involved in neutrino experiments, the RMSGA cross sections compare very well with the RDWIA ones down to remarkably low ejectile kinetic energies of about $200$ MeV. Below this energy, the RMSGA predictions are not realistic due to the underlying approximations, mainly the postulation of linear trajectories and frozen spectator nucleons.

\begin{figure*}[t]
  \begin{center} 	
    \includegraphics[width=15cm]{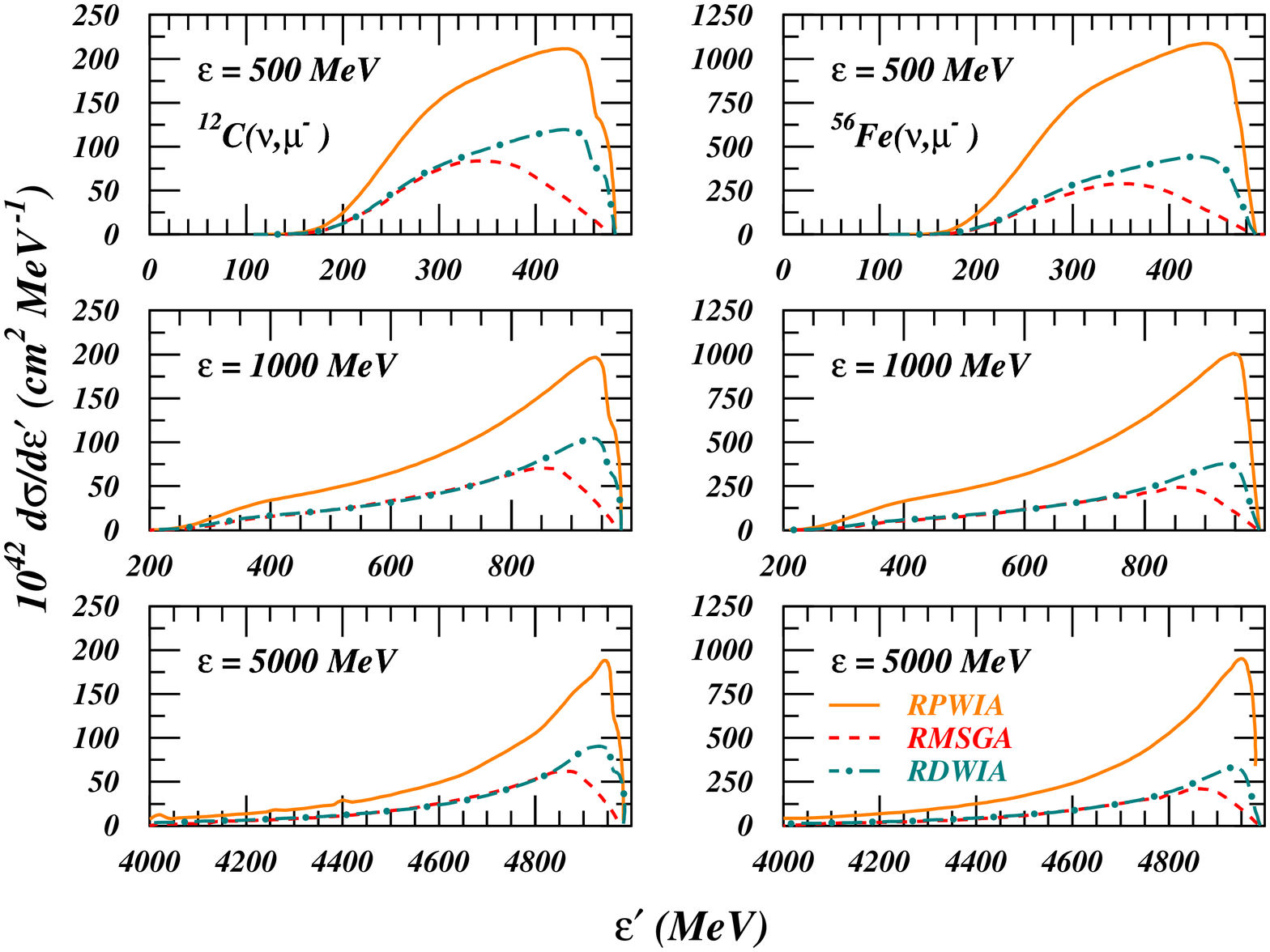}
   \caption{Charged current $^{12}$C$({\nu}_{\mu},{\mu}^-)$ (left
   panels) and $^{56}$Fe$({\nu}_{\mu},{\mu}^-)$ (right panels) cross
   sections as a function of the outgoing lepton energy $\varepsilon'$ at different incoming energies. The labeling is the same as in Fig. \ref{fig.:c12nc}.}
\label{fig.:fe56cc}
  \end{center} 
\end{figure*}

For the sake of completeness, in Fig.~\ref{fig.:fe56cc} we show our predictions for CC $\nu$-nucleus cross sections. The effects of FSI are of the same order as for NC, and very similar results are also obtained within RMSGA and RDWIA down to very low ejectile kinetic energies. 

\subsection{Estimating the effect of FSI mechanisms}
A quantity routinely used to estimate the overall effect of FSI in nucleon-emission processes is the nuclear
transparency. Intuitively, it provides a measure
for the probability that a nucleon of a certain energy - above the
particle-emission threshold - can escape from the nucleus without being subject to any
further interaction. From this 'definition', one can expect that the nuclear transparency is identical for neutrino and electron induced nucleon knockout. Once the nucleon is traversing the nuclear medium, only its energy
is expected to determine the way it propagates. In addition, neutrinos and electrons can be expected to probe equal amounts of bulk and surface parts of the target nucleus.

\begin{figure*}[t]
  \begin{center} 	
    \includegraphics[width=10cm]{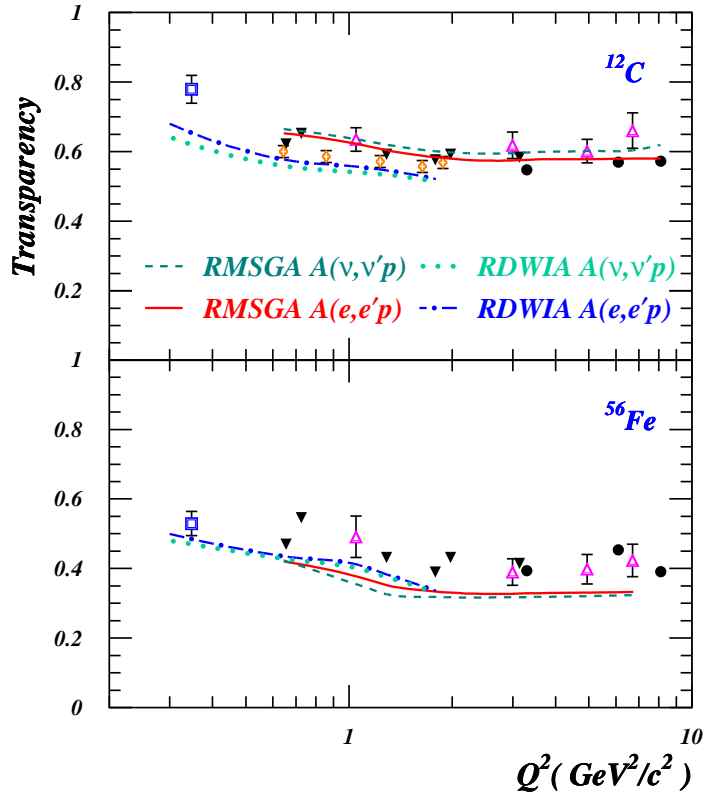}
   \caption{Nuclear transparencies versus $Q^2$ for different nuclei
  in quasi-elastic kinematics. The solid (dot-dashed) lines shows the results of a
  RMSGA (RDWIA) $A(e,e'p)$ calculation \cite{lava04}. The dashed (dotted) lines
  represent the results for $A(\nu,\nu'p)$ within RMSGA (RDWIA). Data points are from Refs.~\cite{Garino92} (open squares),
  \cite{oneill95,makins94} (open triangles), \cite{Garrow2002} (solid
  circles), \cite{Abbott98,Dutta2003} (solid triangles), and \cite{rohe05} (diamonds).}
\label{fig.:neutrinotransp}
  \end{center} 
\end{figure*}

Several investigations of
the nuclear transparency have been carried out using the $A(e,e'p)$
reaction in the QE regime (i.e. the Bjorken variable $x=Q^2/(2M_N\omega)\approx 1$), and data for different nuclei are now available. The nuclear transparency is extracted from the ratio of the measured $A(e,e'p)$ yield to the calculated one using the plane-wave impulse approximation, according to
\begin{equation}
T_{exp}(Q^2)=\frac{\int_Vd^3p_mdE_mY_{exp}(E_m,\vec{p}_m,\vec{k}_f)}{c(A)\int_{V}d^3p_mdE_mY_{PWIA}(E_m,\vec{p}_m)}.
\label{eq:transp}
\end{equation}
The quantity $V$ specifies the experimental phase-space in missing momentum ($p_m$) and energy ($E_m$). The kinematics cuts $|p_m| \leq 300$ MeV/c and $E_m \leq 80$ MeV, in combination with the requirement that  $x\approx 1$, guarantee that the electro-induced proton-emission process is predominantly quasi-elastic. The factor $c(A)$ is introduced to correct in
 a phenomenological way for short-range mechanisms, and is assumed to be moderately target-mass dependent ($c=0.9$ for $^{12}$C, and $c=0.82$ for $^{56}$Fe). It accounts for the fact that short-range correlations move a fraction of the single-particle strength to higher missing energies and momenta and, hence, beyond the ranges covered in the integrations of Eq.~(\ref{eq:transp}). Without going into details, theoretical predictions are obtained in a similar way from the ratio of distorted-wave calculations to plane-wave ones. In Fig.~\ref{fig.:neutrinotransp}, the transparencies predicted by the RMSGA and the RDWIA models are displayed as a function of $Q^2$ for $^{12}$C
and $^{56}$Fe, together with the world data extracted from $A(e,e'p)$. Solid (dot-dashed) lines show the $A(e,e'p)$ results within RMSGA (RDWIA). Details about the calculations can be found in Ref.~\cite{lava04}. The dashed (RMSGA) and dotted (RDWIA) curves correspond to the computed $A(\nu,\nu'p)$ transparencies, obtained using the same procedure as for electron scattering. This procedure includes the computation of RDWIA and RPWIA cross sections at $x \approx 1$, averaged over the same phase space used in Eq.~(\ref{eq:transp}). As can be seen, within each model the neutrino transparencies agree quite well with their electron counterparts. This result clearly illustrates the fact that {\it in our models} the average attenuation effect of the nuclear medium on the emerging nucleon is rather independent of the nature of the leptonic probe.

 Adopting the idea that the nuclear transparency for electrons equals the one for neutrinos,
 the information obtained about nucleon propagation via $A(e,e'p)$ can be used to predict the effects
 of FSI mechanisms in inclusive QE $\nu$-nucleus cross sections. As the transparency is essentially
 the ratio of cross sections including FSI to the ones in the plane-wave limit, this will
 be done by multiplying the RPWIA results for neutrino-nucleus cross sections with
 the measured transparency factors extracted from $A(e,e'p)$. 
In this scenario, the benchmark 
RPWIA neutrino-nucleus cross sections are crucial. It is important to realize that we use 
transparency factors that are confined to $x\approx 1$, while the computation of the 
inclusive neutrino-nucleus cross section include the full phase-space.


\begin{figure*}[t]
  \begin{center} 	
    \includegraphics[width=15cm]{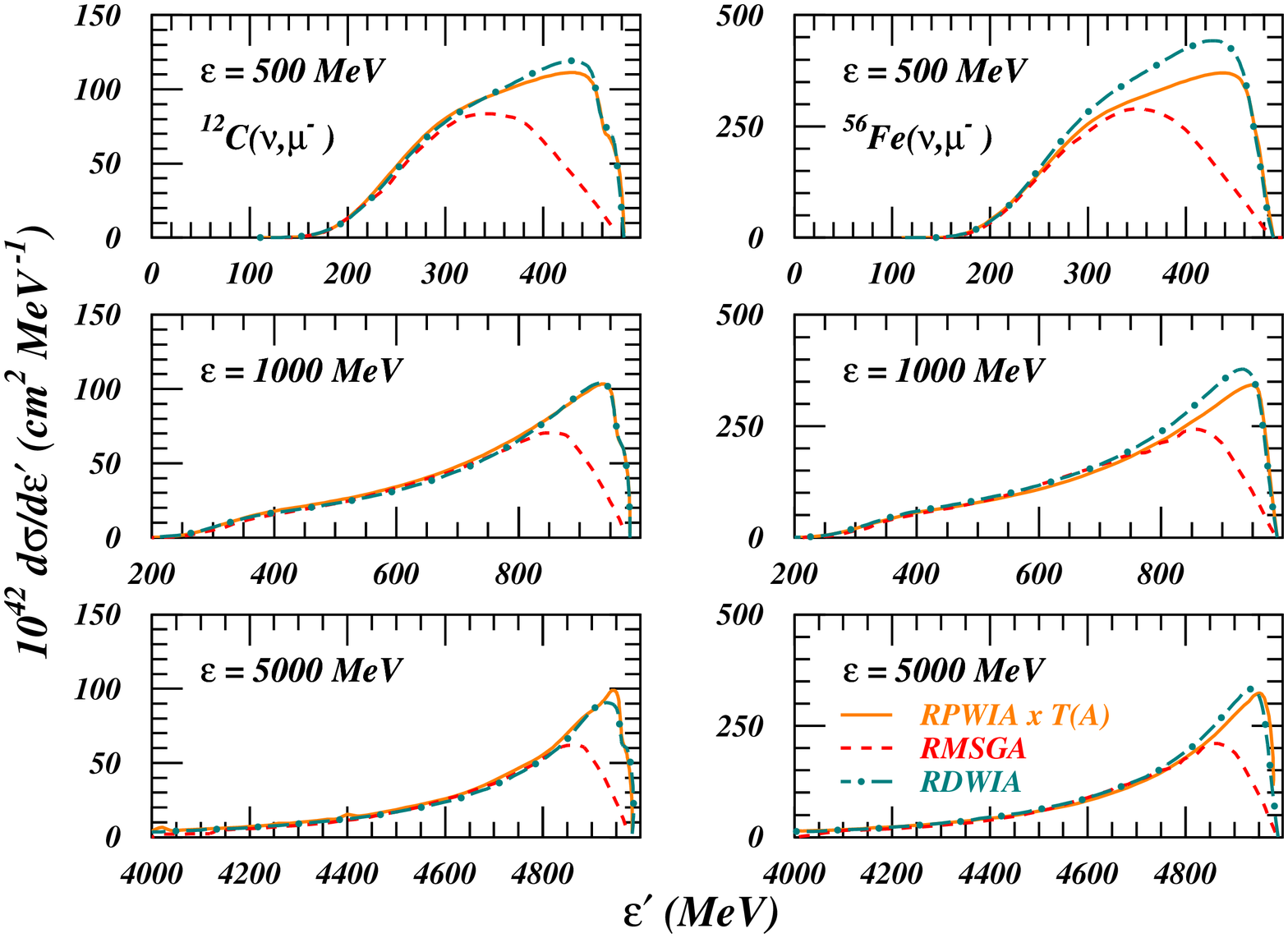}
   \caption{Charged current $^{12}$C$({\nu}_{\mu},{\mu}^-)$ (left
   panels) and $^{56}$Fe$({\nu}_{\mu},{\mu}^-)$ (right panels) cross
   sections as a function of $\varepsilon'$ at different incoming
   energies.  The dashed
   (dot-dashed) lines represent the RMSGA (RDWIA) prediction. The solid lines
   show the RPWIA results, scaled with a transparency factor $T(^{12}C)
   \approx 0.52$ and $T(^{56}Fe) \approx 0.34$.}
\label{fig.:fe56nc}
  \end{center} 
\end{figure*}

In Fig.~\ref{fig.:fe56nc}, the dashed and dot-dashed lines represent the inclusive CC $\nu$-nucleus cross section within RMSGA and RDWIA, respectively. The solid curve
displays our corresponding RPWIA calculation, scaled with a constant factor taken as a representative value for the experimental $A(e,e'p)$ transparency for the nucleus. For $^{12}$C ($^{56}$Fe) we take $T \approx
0.52$ ($ \approx 0.34$).  In extracting these values, we have corrected the measured transparencies from Fig.~\ref{fig.:neutrinotransp} with the factor $c(A)$. A very good agreement is observed between the rescaled RPWIA and the full RDWIA/RMSGA curves in the case of $^{12}$C. This finding supports the idea that a simple scaling of the RPWIA results with a transparency factor obtained from electron scattering data allows one to reliably estimate the FSI effects for the quasi-elastic contribution to the inclusive neutrino cross section. The fact that for $^{56}$Fe the agreement is less satisfactory reflects the fact that our models slightly underestimate the $^{56}$Fe transparency data.



\begin{figure*}[t]
  \begin{center} 	
    \includegraphics[width=15cm]{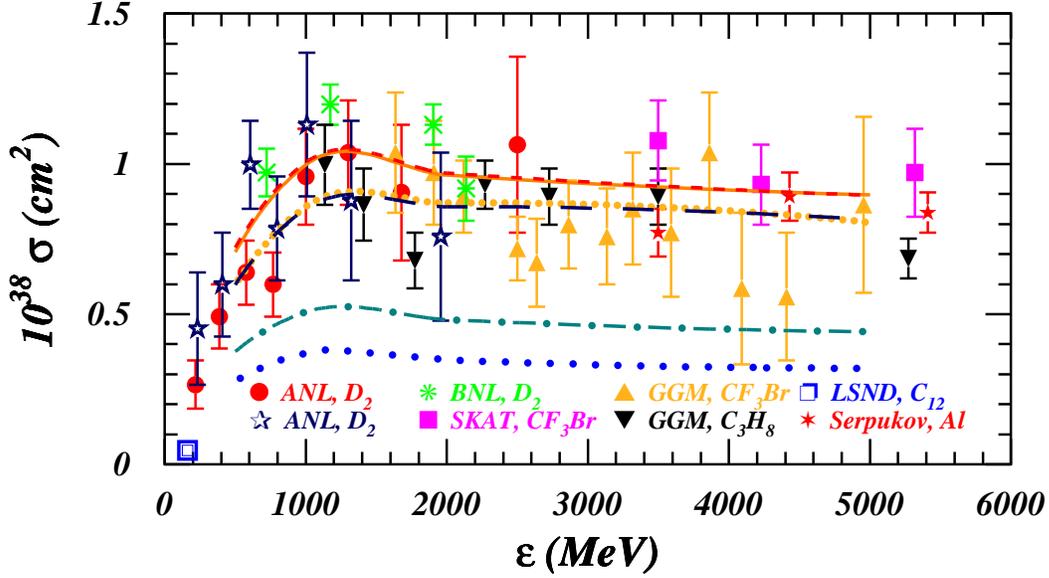}
   \caption{Total CC $(\nu_{\mu},\mu^-)$ neutrino cross sections
   as a function of the incoming neutrino energy. The solid (dashed) line shows
   the RPWIA calculations on $^{12}$C ($^{56}$Fe). The dot-dashed
   (long-dotted) curves implement the effect of FSI on $^{12}$C ($^{56}$Fe)
   within RDWIA. The short-dotted (long-dashed) show the RMF curve for $^{12}$C ($^{56}$Fe). All results are scaled with the number of neutrons in the target. Data
    points are from Refs.~\cite{barish,baker,mann,brunner,pohl,auerbach,belikov,bonetti}.}
\label{fig.:totalcross}
  \end{center} 
\end{figure*}

Finally, Fig. \ref{fig.:totalcross} displays the total cross section
 ($\sigma=\int d\varepsilon'(d\sigma/d\varepsilon')$) for $^{12}$C$(\nu_{\mu},\mu^-)$ and
 $^{56}$Fe$(\nu_{\mu},\mu^-)$ reactions, scaled with the number of neutrons in the target. Results
 are shown within RPWIA and RDWIA using a complex optical potential. The figure clearly shows that
 the difference between RPWIA and RDWIA cross sections is approximately given by the experimental
 transparency factor extracted from $A(e,e'p)$ at QE kinematics. Furthermore, other important features
 can be extracted from this figure. First, the RPWIA cross sections scale with the target mass-number. 
In this way, when RPWIA cross sections are required for a heavy nucleus, a very good approximation 
consists in multiplying this cross section per nucleon by its mass number. Second, the cross sections do not
 appreciably change from neutrino energies above 2 GeV, i.e. the cross sections saturate at high incoming 
neutrino energies. To finish with, we compare our relativistic calculations with data from various experiments.  The RPWIA calculations give a fair account
of the neutrino-energy and magnitude of the data. The RPWIA is
confined to single-nucleon knockout thereby not including final-state
interaction effects. The RDWIA calculations, on the other hand,
including FSI effects via the introduction of an optical potential,
considerably underestimate the data. The results contained in Fig. \ref{fig.:totalcross}
indicate that at least $50\%$ of the measured $(\nu_{\mu},\mu ^{-})$
strength can be attributed to single-step ("elastic") nucleon knockout
to missing energies below 80 MeV in the residual A-1 nucleus.  The
remaining fraction of about $50\%$ could be attributed to multi-nucleon
knockout, pion production with subsequent reabsorption, single-nucleon knockout to more complex
states, ... .  Adding all these contributions would move the
calculations closer to the data.

It has been suggested~\cite{serot86, kim3, caballero05} that the importance of these missing channels can be
estimated in a model in which both the single-particle bound-states
and scattering states are computed in a real mean-field potential
obtained in the Hartree approximation to the $\sigma-
\omega$ model. Such an extreme mean-field model (here coined as RMF) involves no imaginary
potential and accordingly the loss of single-nucleon knockout strength
into "inelastic" nucleon-knockout channels is effectively
reintroduced. As can be appreciated from Fig. \ref{fig.:totalcross}, the presence of a real
potential reduces the RPWIA single-nucleon knockout strength, but the
global reduction with respect to RPWIA is much smaller compared to the $50-60\%$ in the RDWIA/RMSGA frameworks.  It is remarkable that the
energy dependence of the cross section is identical in all three
relativistic frameworks adopted here. The magnitude, on the other
hand, depends strongly on the model used to account for FSI
mechanisms.  This reinforces our suggestion that one could use RPWIA
results to predict the "elastic" single-nucleon knockout contribution
to inclusive neutrino cross sections, provided that they are rescaled
with a transparency factor extracted from $A(e,e'p)$ data.

\section{Conclusions}
\label{sec:conclusions}

We have employed two relativistic models, RMSGA and RDWIA, to study NC and CC quasi-elastic neutrino-nucleus scattering. Results have been presented for carbon and iron targets, covering a wide range of neutrino energies. Within RPWIA both models provide nearly identical results, which deviate from existing RPWIA predictions. The fact that two independently developed codes that adopt very different numerical techniques agree in this limit, together with the fact that our RPWIA predictions approach the RFG model at high energies, give us confidence that our RPWIA calculations serve as benchmark results. We subsequently computed the effects of FSI mechanisms within the RMSGA and RDWIA models. The two ways of dealing with FSI are consistent down to remarkably low outgoing nucleon kinetic energies of about 200 MeV. FSI produce a large reduction of the cross sections, that increases with the mass number of the target nucleus. Finally, we have illustrated that the nuclear transparencies extracted from $A(e,e'p)$ measurements can be used to estimate the effect of FSI mechanisms on the elastic single-nucleon knockout contribution to the inclusive neutrino-nucleus cross sections. Extensions of our models include the implementation of pion production and delta resonance. Work in this direction is in progress.


%


\end{document}